\documentclass[12pt]{article}

\usepackage{scicite}
\usepackage{graphicx}
\usepackage{times}
\usepackage{amsmath, amssymb, amsfonts, amsthm, fouriernc}
\usepackage[nodisplayskipstretch]{setspace}

\newenvironment{sciabstract}{%
\begin{quote} \bf}
{\end{quote}}

\title{Tipping in complex systems under fast variations of parameters}

\author
{Induja Pavithran,$^{1,2}$ P. R. Midhun,$^{2}$ R. I. Sujith$^{2\ast}$\\
\\
\normalsize{$^{1}$Department of Physics, IIT Madras, Chennai, India}\\
\normalsize{Department of Aerospace Engineering, IIT Madras, Chennai, India}\\
\\
\normalsize{$^\ast$To whom correspondence should be addressed; E-mail:  sujith@iitm.ac.in}
}

\date{}

\begin{document} 


\baselineskip24pt


\maketitle


\begin{sciabstract}
 Sudden transitions in the state of a system are often undesirable in natural and human-made systems. Such transitions under fast variation of system parameters are called rate-induced tipping. We experimentally demonstrate rate-induced tipping in a real-world complex system and decipher its mechanism. There is a critical rate of change of parameter above which the system undergoes tipping. We show that another system parameter, not under our control, changes simultaneously at a different rate, and the competition between the effects of that parameter and the control parameter determines if and when tipping occurs. Motivated by the experiments, we use a nonlinear oscillator model exhibiting Hopf bifurcation to generalize this tipping to complex systems in which slow and fast parameters compete to determine the system dynamics.
\end{sciabstract}



\section*{Introduction}

Rising levels of greenhouse gases in the atmosphere continue to heat our planet, which is rapidly approaching a critical climate change leading to unanticipated catastrophic consequences \cite{lenton2019climate,lovejoy2019amazon}. Climate change can affect ecosystems and trigger abrupt transitions exhibiting sudden changes in their state \cite{carpenter1999management,scheffer2001catastrophic}. 
Such transitions include occurrence of algal blooms in lake ecosystems, epileptic seizures, asthma attacks, migraines, extinction of species and desertification \cite{venegas2005self,litt2001epileptic,ortiz2020detecting}. A sudden transition to a contrasting state due to a gradual change in the system parameter is generally known as `tipping'\cite{lenton2008tipping}. In real-world complex systems, these transitions are often `not easily reversible' and have prolonged consequences that even can result in the collapse of the entire system.
Here, complex systems refer to systems comprising many interacting subsystems whose interactions can lead to emergent phenomena. The positive feedback mechanisms are found to be the basic ingredients of tipping in complex systems\cite{scheffer2012anticipating,angeli2004detection}.

The use of bifurcation theory helped understand the problem of unexpected tipping in many systems \cite{kuznetsov1998elements,thompson2011predicting}. Whenever the tipping is induced by a bifurcation, the stability margin can be identified by estimating the stability of the equilibrium states. In such cases, the tipping point is nearly the same as the bifurcation point. However, the tipping point can deviate from the bifurcation point due to the presence of random fluctuations (noise) in the system or time-dependent variation of the  system parameters \cite{baer1989slow,unni2019interplay}. The system relaxes back to the stable equilibrium for small perturbations due to noise. For sufficiently high noise intensity, there is a possibility of escaping from the basin of attraction, before reaching the expected tipping point (N-tipping) \cite{ashwin2012tipping,ritchie2017probability,ditlevsen2010tipping}. Further, parameters varying continuously at a finite rate can delay tipping; this is commonly known as rate-delayed tipping or slow passage through bifurcation \cite{baer1989slow}. On the other hand, instances of advanced tipping are also reported \cite{suchithra2020rate,manikandan2020rate}.   

Many real-world systems are non-autonomous. In natural systems, the parameters often vary continuously on their own, whereas we vary them intentionally in engineering systems. Throttling in aircraft engines is an example of a situation where we change the parameter continuously. In contrast, parameters in the climate system such as global temperature or planetary albedo vary on their own. Therefore, one should consider the rate at which the parameter varies while studying critical transitions in real systems. 
Fast rates of change of parameters introduce interesting effects on tipping, including unexpected tipping even without having an underlying bifurcation. 

Ashwin et al.\cite{ashwin2012tipping} classified this category of tipping as rate induced tipping (R-tipping), where a slow variation of a parameter does not show any tipping. Only fast variations of the parameter (faster than a critical rate) lead to tipping; such tipping does not need any change in the stability of equilibrium states.
They defined R-tipping as a condition where the system fails to track the continuously changing quasi-steady attractors and tips to an alternative stable state. While the quasi-steady or sufficiently slow variation of parameter does not result in any bifurcation, a continuous variation at a rate faster than the critical rate results in tipping. In other words, the dynamical system, $\dot{x} = f(x,a)$, does not exhibit bifurcation upon varying the parameter $a$; however, it can undergo transition upon changing $r$, where $\dot{a} =r$. Note that $a$ is not a bifurcation parameter here.
Recently, Tony et al.\cite{tony2017experimental} discovered a different mechanism of R-tipping called preconditioned R-tipping in an experimental thermoacoustic system and illustrated it in a model exhibiting subcritical Hopf bifurcation. They varied the bifurcation parameter itself continuously at a finite rate within the bistable region, and R-tipping was achieved by preconditioning the system with a suitably high initial perturbation. This initial perturbation (less than the amplitude of the unstable limit cycle) alone is not sufficient to induce tipping in the bistable region; it will decay, and the dynamics will approach the fixed point. Fast rates of change of the parameter above a critical rate help the system cross the unstable limit cycle, and directly approach a stable limit cycle instead of decaying to a fixed point. 

In all the above scenarios, the variation of a system parameter has to exceed a threshold rate for a runaway change to occur, wherein the system abruptly leaves an attractor. The resulting change can be reversible as well, as described by Wieczorek et al. \cite{wieczorek2011excitability}.
Their work on excitable slow-fast systems showed the possibility of rate dependency and a different type of tipping in a climate system model. A reversible type of R-tipping was observed, where the system can be excited, with a ramped parameter, from the existing attractor and return to it repeatedly. They used the climate-carbon cycle model with global warming to explain a potential climate tipping point known as the compost-bomb instability (an explosive release of soil carbon into the atmosphere).


The dangerous sudden tipping due to fast variation of parameters may come with no early warning. Most of the generic early warning signals work by capturing the signatures of the underlying bifurcation \cite{carpenter2006rising,van2007slow}. However, R-tipping need not be accompanied by a change of stability of the system (bifurcation); as a result, the occurrence of R-tipping is challenging to predict. Therefore, we need to understand the mechanisms behind them to predict such transitions, especially in real-world complex systems.

The studies explaining the mechanism of R-tipping are either theoretical or based on models \cite{wieczorek2011excitability,ashwin2012tipping,luke2011soil,ritchie2017probability} and do no have any experimental support. At the same time, experimental observations of R-tipping \cite{manikandan2020rate} lack the understanding of the underlying mechanism. 
The existing theories do not explain these results, as accessing all system parameters is impossible to achieve. To address this issue, we perform experiments in a real-world complex system, a turbulent thermoacoustic system, by measuring another slow varying parameter to identify the mechanism.
We experimentally show a mechanism of R-tipping in complex systems with positive feedback. 

\section*{R-tipping in a turbulent thermoacoustic system}

We study R-tipping in an engineering system - a combustion-based power generating system.
Examples are power-producing gas turbine engines, rocket engines and jet engines used for aviation. In such combustion systems, the perturbations in the unsteady flow field causes heat release rate fluctuations from the flame resulting in sound production which, in turn, perturbs the flame. In a confined environment, the sound waves get reflected from the walls of the combustor and interact with the flame. Originating from the nonlinear interactions between these sound waves, the hydrodynamic field, and the flame inside the combustor, the thermoacoustic system exhibits rich dynamics. A turbulent thermoacoustic system is a complex system that exhibits dynamical transitions and emergent dynamics depending on the interactions between the subsystems. Under certain operating conditions, the interactions between these subsystems establish a positive feedback mechanism. This positive feedback mechanism drives the system towards a self-organized state known as thermoacoustic instability\cite{sujith2020complex}.

Industrial combustors designed to operate under stable operating conditions produce hazardous emissions such as NOx, CO and SOx as by-products of burning fossil fuels. This emission problem is often tackled by operating them at ‘greener' conditions, i.e., fuel-lean conditions which produce lower emissions. However, operating the engine in such a fuel-lean condition makes it susceptible to the phenomenon of thermoacoustic instability, which manifests as ruinously high amplitude pressure oscillations. The transition to thermoacoustic instability is a critical transition that is undesirable in all combustion systems. The spontaneous emergence of high amplitude pressure oscillations in gas turbine engines and rocket motors is a persistent challenge faced by the propulsion and power industry\cite{juniper2018sensitivity,sujith2021thermoacoustic}. The phenomenon of thermoacoustic instability can cause catastrophic damage to the system through severe vibrations leading to structural failure, fatigue, failure of thermal protection systems, failure of navigation and control systems, and reduced life span of the engine\cite{lieuwen2005combustion}. The problems of thermoacoustic instability have even led to the failure of space missions\cite{fisher2009remembering}. 

The present study exploring R-tipping is relevant especially in thermoacoustic systems, because,  
the operating conditions which appear to be stable under quasi-static or slow variation of parameters may exhibit thermoacoustic instability. Hence, rate-induced transitions result in the unexpected occurrence of thermoacoustic instability and are even more dangerous as it is hard to detect the transition during quasi-steady experiments. Hence, we consider thermoacoustic system as an example of a complex system where investigating possibility and mechanism of R-tipping is highly relevant.

\begin{figure}
\centering
\includegraphics[width= 0.65\linewidth]{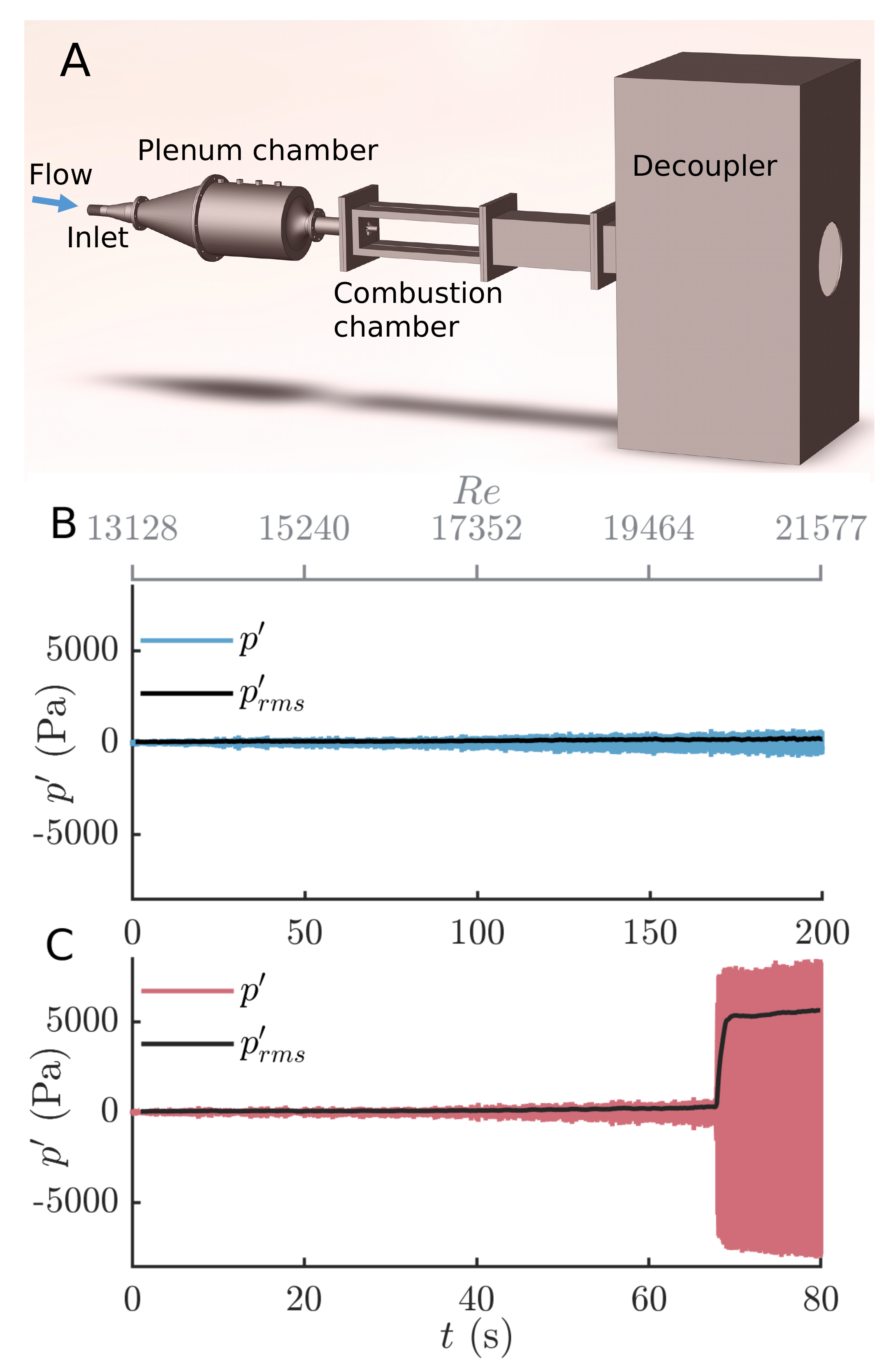}
\caption{(A) Schematic of the laboratory-scale turbulent combustor used for this study. A backward-facing step combustor with a combustion chamber having a cross-section of 90 mm $\times$ 90 mm and a length of 800 mm is used. The air flow enters through the inlet to the plenum chamber. We vary the $Re$ of the flow continuously.
Time series of pressure oscillations acquired for (B) a slow variation ($dRe/dt = 42\ s^{-1}$) and (C) a fast variation of Reynolds number ($dRe/dt = 106\ s^{-1}$). Initially, we observe a state of low amplitude aperiodic fluctuations for low values of $Re$ in both cases. Then, the system transitions to a state of periodic oscillations with very high amplitude ($\sim$8000 Pa) for the fast variation of $Re$. Even though we are varying $Re$ in the same range, we do not observe a high amplitude limit cycle for the slow variation of $Re$.}
\label{fig:fig3}
\end{figure}

We explain the experiments briefly here; a detailed description of the experimental setup and measurements are given in Methods. The turbulent thermoacoustic system comprises a backward facing step combustor, a plenum chamber, and a decoupler (Fig.~\ref{fig:fig3}A).
The air flow enters through the inlet to the plenum chamber and the fuel (liquefied petroleum gas) is injected upstream of the combustion chamber. After ignition, once the flame is stabilized in the combustor, we vary the mass flow rate of air, which, in turn, varies the Reynolds number ($Re$). $Re$ is considered as the control parameter in this study and is calculated as $Re = 4\dot{m}/\pi \mu D_0$, where $\dot{m}$ is the total mass flow rate of the air-fuel mixture, $\mu$ is the dynamic viscosity of the mixture and $D_0$ is the diameter of the circular duct just before the combustion chamber. 
In the present study, we vary $Re$ continuously from $1.3\times10^4$ to $2.2\times10^4$. 
To study the dynamical transitions in the system, we measure the pressure fluctuations inside the combustor  and the wall temperature of the combustion chamber.

We perform experiments by varying $Re$ at different rates (from $dRe/dt$ = 35.2 $s^{-1}$ to 140.8 $s^{-1}$). We keep all the operating conditions the same, except the rate of variation of the control parameter. The rate of change of the Reynolds number, $dRe/dt$, is kept constant for a particular experiment. First, we present an analysis of the data acquired from two trials of experiments with two different values of $dRe/dt$. One experiment is performed at a relatively slow rate ($dRe/dt = 42\ s^{-1}$), and the total duration of the experiment is 200 s (Fig.~\ref{fig:fig3}B). The system remains in a state of low amplitude aperiodic fluctuations throughout the entire range of Reynolds numbers. Next, we present another data for an experiment conducted at a faster rate ($dRe/dt = 106\ s^{-1}$) where the Reynolds number is varied in the same range, but for a duration of 80 s (Fig.~\ref{fig:fig3}C). Here, as the Reynolds number varies, the system exhibits a transition from low amplitude aperiodic fluctuations to high amplitude limit cycle oscillations. We observe a sudden jump from a low amplitude state to a high amplitude periodic oscillatory state. During the state of low-amplitude fluctuations, the amplitude spectrum shows a wide band of frequencies, and after the transition, it becomes a narrow peak centered around 190 Hz (not shown here).

\begin{figure}[h]
\centering
\includegraphics[width= 0.943\linewidth]{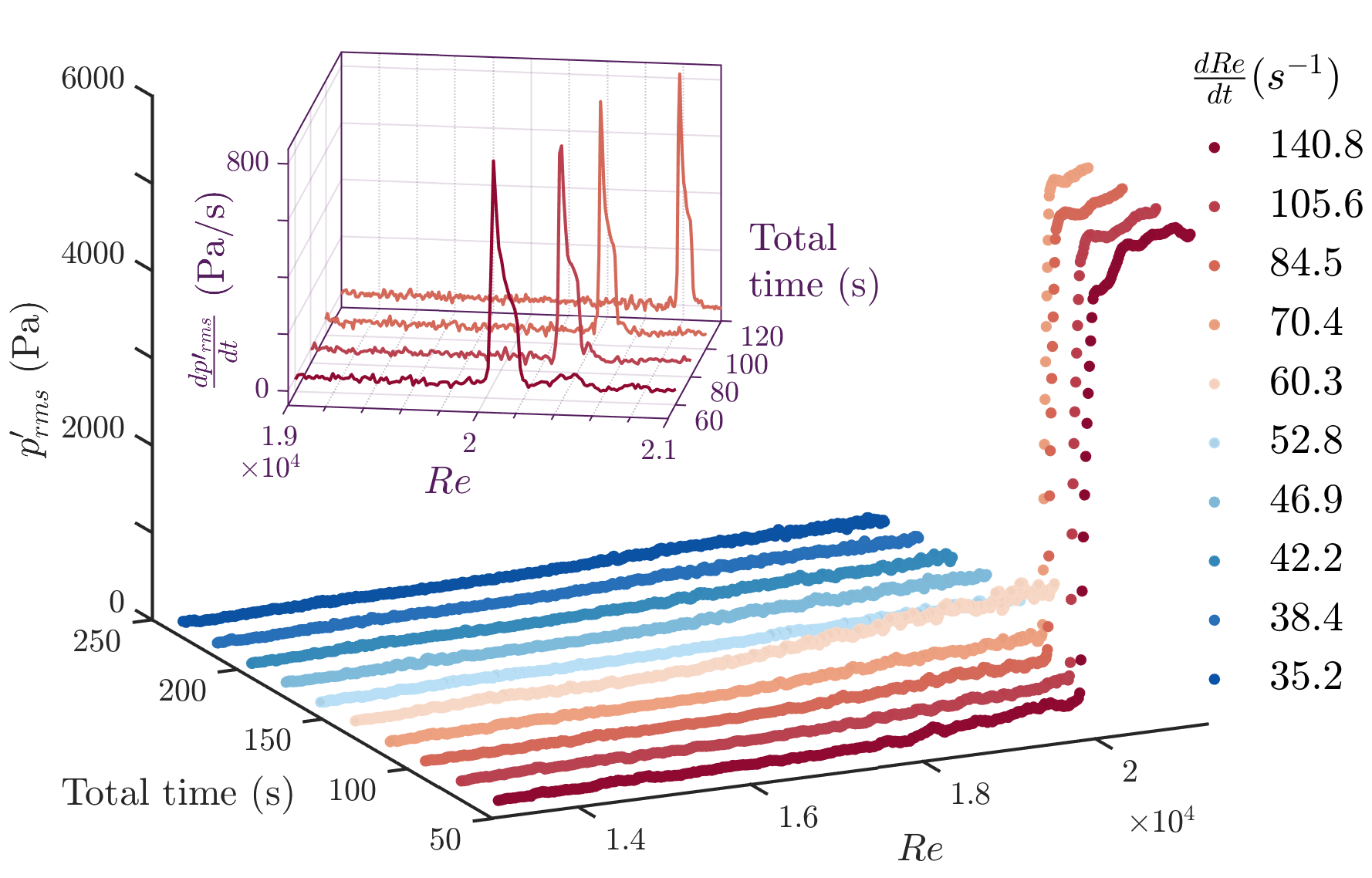}
\caption{Bifurcation diagram for experiments performed at different $dRe/dt$. We vary $Re$ in the same range at different rates, thereby having different duration for the experiments. For each experiment, we plot the rms of pressure oscillations as a function of $Re$. We observe rate-induced tipping to high amplitude limit cycle oscillations for faster variations of the parameter ($dRe/dt>60.3$). Slower rates of variation below a critical rate do not lead to a jump in $p^\prime_{rms}$. The inset figure shows the rate of change of $p^\prime_{rms}$, wherein the maximum increase in amplitude appears as a peak. This helps to define a tipping point when control parameters vary continuously. The advanced onset of thermoacoustic instability with fast rates of change of parameter can be clearly observed here. }\label{fig:fig4}
\end{figure}

To analyze the experiments performed at different $dRe/dt$, the root mean square (rms) of acoustic pressure fluctuations, calculated for a moving window of 1 s and overlap of 0.9 s, is plotted as a function of $Re$ for different values of $dRe/dt$ (Fig.~\ref{fig:fig4}).
We do not observe tipping for slow rates of variation of $Re$ ($dRe/dt\leq60.3\ s^{-1}$), whereas, fast rates of variation of $Re$ ($dRe/dt>60.3\ s^{-1}$) result in a transition to thermoacoustic instability. We observe two different dynamics at rates lower and higher than a critical rate, for the same range of control parameter values. $dRe/dt \sim 60.3\ s^{-1}$ is the `experimentally observed' critical rate in this case. This idea of the critical rate will be clearer when we discuss the model.

The inset in Fig.~\ref{fig:fig4} shows the rate of change of $p^\prime_{rms}$ which helps to identify the point of maximum growth in the amplitude of pressure fluctuations. Such a point of maximum change in $p^\prime_{rms}$ is identified as the tipping point \cite{pavithran2021effect}. This plot is drawn considering only the cases where there is tipping (i.e., for rates faster than the critical rate, $dRe/dt>60.3\ s^{-1}$). An interesting observation is that there is an advancement in the tipping point with the increase in the rate change of parameter, contrary to the rate-dependent tipping-delay observed in earlier studies \cite{baer1989slow,pavithran2021effect,tandon2020bursting}. The peak in the plot shifts towards lower values of $Re$ with an increase in the rate of change of parameter; i.e., the tipping point advances for faster rates. Although advanced tipping has been reported earlier in various complex systems \cite{suchithra2020rate,manikandan2020rate}, to date, there is no physical explanation for such advancement of tipping with fast rates of variation of parameters. To identify when to expect advanced tipping (contrary to delayed tipping) in non-autonomous systems, we proceed to investigate the mechanism behind the rate induced tipping observed in this thermoacoustic system and then generalize it by illustrating it in a mathematical model.

\section*{The mechanism of R-tipping}
 
We investigate the mechanism behind rate-induced tipping to high amplitude limit cycle oscillations observed in our thermoacoustic system. We examine the evolution of the wall temperature of the combustor to understand the dynamics during the transition. The walls of the combustor get heated up gradually during the experiment, and hence the wall temperature varies continuously at a finite rate. Therefore, the wall temperature is another parameter that varies simultaneously with the control parameter at a different rate. Thus, we have a non-autonomous dynamical system.

\begin{figure}[h!]
\centering
\includegraphics[width=0.55\linewidth]{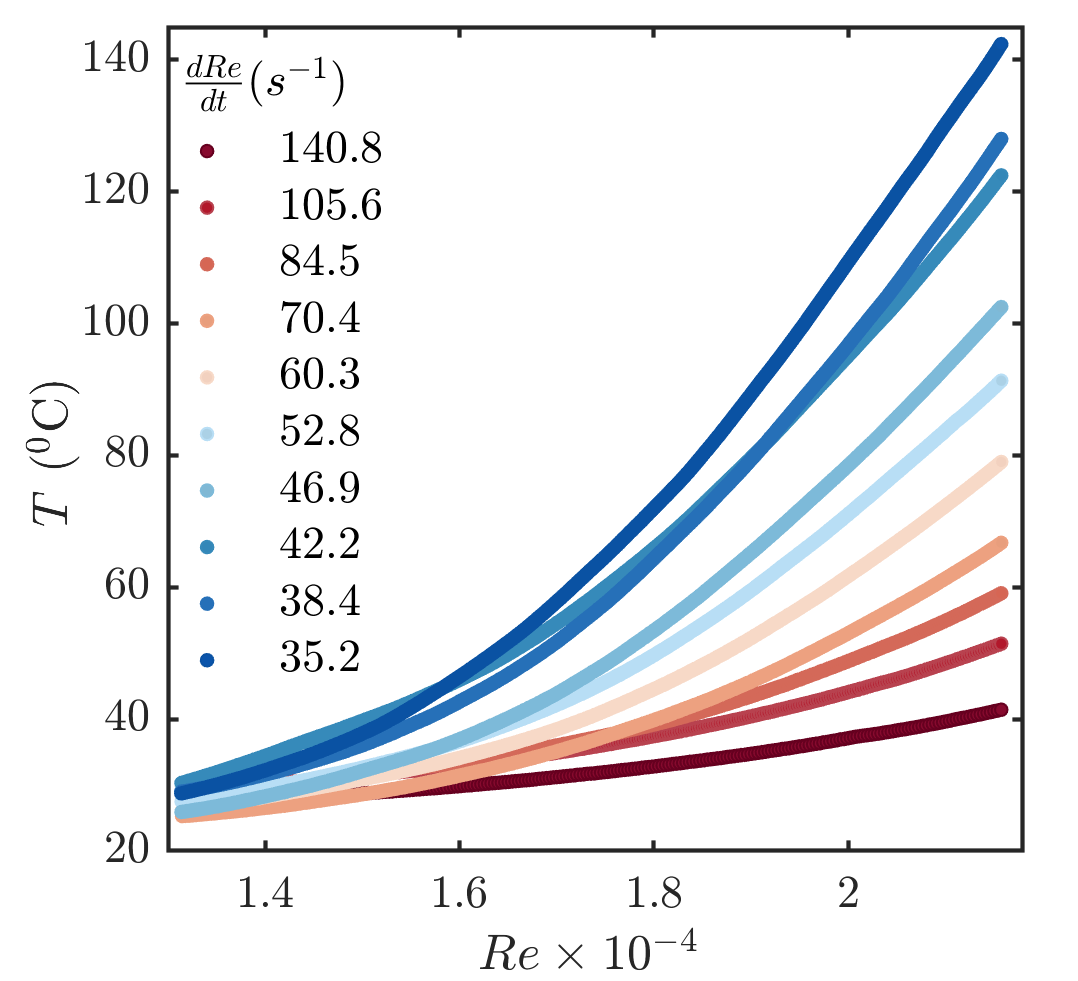}
\caption{Variation of wall temperature measured in time as a function of the control parameter. The dark blue curves, which are increasing up to high temperatures, represent slow rates. In contrast, the faster variations of $Re$ result in comparatively lower values of wall temperature at a given value of $Re$ (dark red). Even though the increase in wall temperature is gradual in time, due to the difference in the duration of the experiments, we obtain significantly different wall temperatures for a given $Re$ for experiments at different $dRe/dt$. Hence, different paths are obtained in the [$Re,\ T$] parameter plane.
Note that each curve in the figure corresponds to experiments of different duration.}
\label{fig:fig5}
\end{figure}

Figure~\ref{fig:fig5} shows the variation of wall temperature with change in $Re$. Note that we start all the experiments at room temperature (25$^\circ C \pm 1^\circ C$). Then, the control parameter is varied after waiting 5 s from the ignition. Each of the curves in Fig.~\ref{fig:fig5} corresponds to a different duration of the experiment; therefore, higher values of wall temperatures are attained in the experiments having a longer duration corresponding to slow variations in $Re$. On the other hand, the increase in wall temperature is relatively small for fast variations in $Re$. 
Thus, the wall temperatures for a given $Re$ are different for slow and fast rates; it is relatively higher for slower rates of change of $Re$ compared to that of faster rates.

Therefore, in this 2-parameter plot, the system tracks different directions depending on the rate of change of $Re$ and reaches a different set of parameters. Thus, different dynamics can be expected for these cases, as each set of parameters can drive the system towards different attractors. Note that we do not discuss quasi-static variation in the experiments; we have to keep parameters constant for asymptotically long times, which is not possible in this case as the wall temperature is changing continuously.

Most importantly, the system did not undergo any transition to limit cycle oscillations for slow rates with relatively higher wall temperatures. As tipping is not observed for higher wall temperatures, we examine the effect of wall temperature on acoustic damping.
\begin{figure}[h!]
\centering
\includegraphics[width=0.55\linewidth]{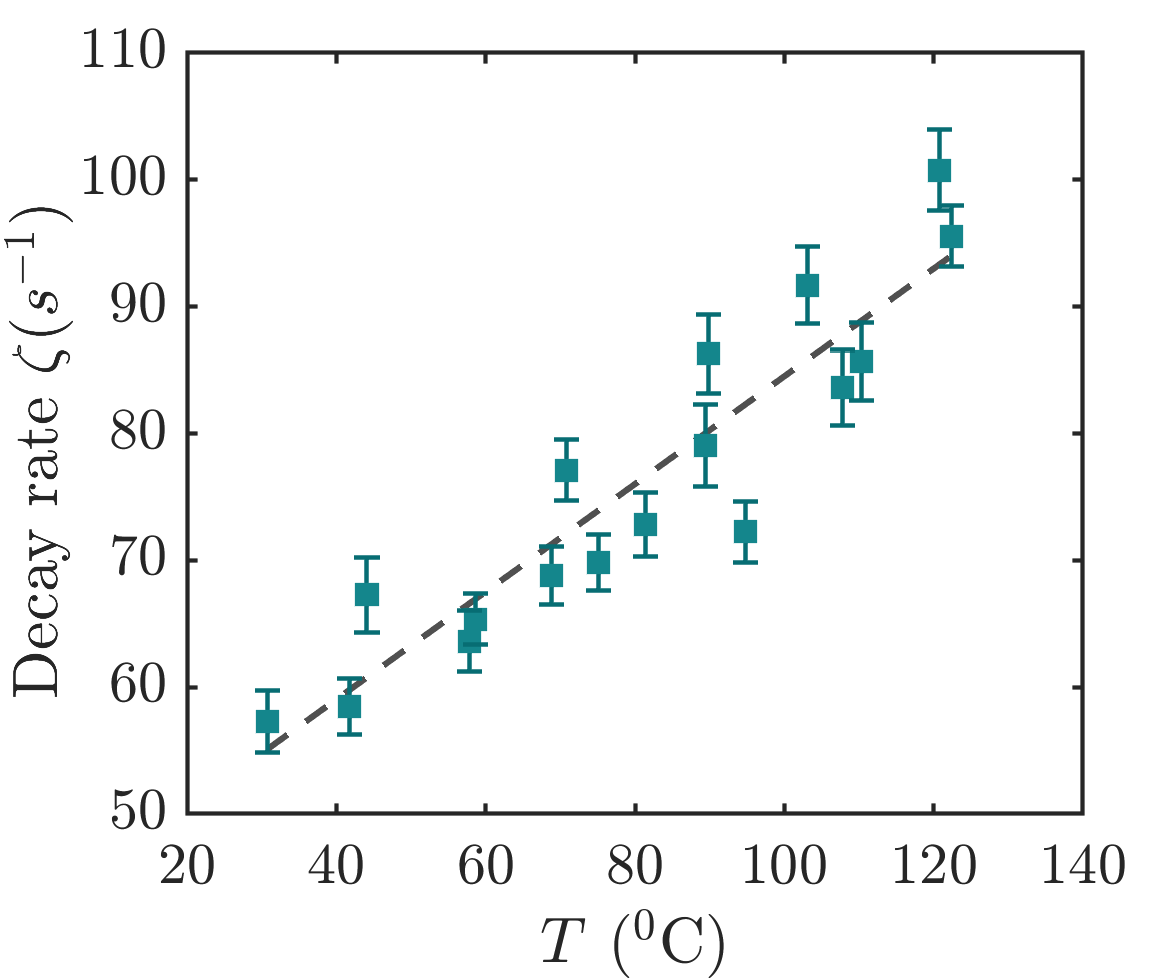}
\caption{To explore the dependency of acoustic damping on the wall temperature, we measure the decay rate of the envelope of acoustic pressure oscillations, $\zeta$, as a function of the wall temperature. We conduct a set of experiments wherein the combustion is stopped abruptly by cutting off the fuel supply during the state of periodic oscillations. We fit an exponential function to the envelope of oscillations \cite{culick2006unsteady,perry1970investigations} to measure the rate of decay of oscillations at different wall temperatures.
We observe that the decay rate (a measure of acoustic damping) increases with increasing wall temperature. We fit a straight line (slope = $0.42 \pm 0.1$) to obtain an empirical relation between the decay rate and the wall temperature. 
}
\label{fig:fig6}
\end{figure}
Towards this purpose, we measure the decay rate ($\zeta$) of acoustic pressure oscillations, which is a measure of acoustic damping, at different wall temperatures. The details of estimation of the decay rate are provided in Methods. 
The plot of the decay rate as a function of the wall temperature (Fig.~\ref{fig:fig6}) shows that the acoustic damping increases with increasing wall temperature, depicting a clear linear relation between them. Interestingly, this type of relation between wall temperature and acoustic damping has not been reported till date in turbulent thermoacoustic systems to the best of our knowledge, and the physical mechanisms behind such a dependency of acoustic damping on temperature need further investigation. An increase in $Re$ corresponds to an increase in acoustic driving in time, and the acoustic damping increases simultaneously at a rate that we do not have any control over.

Although we are varying a single parameter during the experiment, there is another ``hidden'' parameter continuously changing on its own, at a different rate. 
We vary $Re$ continuously to drive the system towards thermoacoustic instability. Concurrently, the continuously increasing wall temperature leads to an increase in the acoustic damping in the system. The transition to thermoacoustic instability occurs when the thermoacoustic driving overcomes the acoustic damping in the system. 
As the walls of the combustor are heated only for a shorter duration for experiments corresponding to fast variation of $Re$, the wall temperatures attained are relatively lower, leading to lower levels of acoustic damping. Therefore, it is easy to overcome acoustic damping at a comparatively lower level of thermoacoustic driving, resulting in an advanced onset of tipping for faster variations of $Re$, as observed in our experiments (see Fig.~\ref{fig:fig4}). 


Following the discussion above, it is clear that, in experiments, the driving matches damping at lower $Re$ for relatively faster rates compared to a critical rate. At the critical rate, both driving and damping increase at the same rate, and they do not ever match. For slightly faster rates compared to the critical rate, driving needs to increase to a very high value to match damping, which may not be feasible to attain in experiments, because we can vary parameters only within a range. In our case, all the experiments are performed within the same range of $Re$ from 13,000 to a maximum value of 21,500. We increase $Re$ by increasing the flow rate of air, whereas the flow rate of fuel is kept constant. If we continue to increase the flow rate of air after a limit, to a fuel-lean condition, the flame becomes more sensitive to flow perturbations. At high flow velocities, eventually, the flame loses its ability to anchor itself inside the combustor and blows out \cite{shanbhogue2009lean,nair2007near}. Since we vary the control parameter within a range, there is a possibility that the actual critical rate is not captured in the experiments and the experimentally observed value may be slightly higher than the actual.

Next, we illustrate and generalize the mechanism of R-tipping discussed using a system-independent nonlinear oscillator model with two time-varying parameters.

\section*{R-tipping in noisy Hopf bifurcation model}
\begin{figure}[h!]
\fbox{
\begin{minipage}[h]{0.98\linewidth}
\setstretch{1.3}
\textbf{Box 1. Hopf bifurcation model.}\\
A simple model of a nonlinear oscillator (a Van der Pol oscillator with higher order terms) exhibits subcritical Hopf bifurcation \cite{ananthkrishnan1998application,noiray2017linear}. 
\begin{equation}
\centering
\ddot{\eta}+\alpha \dot{\eta}+\omega^{2} \eta=\dot{\eta}\left(\beta+\kappa \eta^{2}-\gamma \eta^{4}\right)+\xi \hspace{8.8cm} (5) \nonumber
\end{equation}
Here, $\alpha$ and $\beta$ are the linear damping and driving terms, respectively. The variations in $\alpha$ and $\beta$ can be considered analogous to varying wall temperature and $Re$ in the experiments. Additive white noise $\xi$ with intensity $\Gamma$ and autocorrelation $<\xi \xi_{\tau}> = \Gamma \delta (\tau)$ is added to represent the inherent fluctuations in the system variables. The values of the parameters $\omega$, $\gamma$, $\kappa$ and $\Gamma$ are kept constant ($\omega$ = 2$\pi \times$ 120 rad/s, $\beta$ = 50 rad/s, $\gamma$ = 0.7, $\kappa$ = 9, $\Gamma = 10^5$), following Noiray \cite{noiray2017linear}. The linear damping ($\alpha$) and driving ($\beta$) are varied. Whenever $\alpha$ is greater than $\beta$, any perturbation decays to the fixed point state ($\eta$ = 0) and for $\beta > \alpha$ the system exhibits limit cycle oscillations. The fixed point state appears as low-amplitude oscillations in $\eta$ due the presence of noise. The transition from the state of low amplitude fluctuations to high amplitude limit cycle oscillations occurs when the value of $\beta$ exceeds $\alpha$. The deterministic bifurcation curve ($\Gamma = 0$) is shown for $\alpha = 60$ and $\beta$ as the control parameter.\\

\centering
\includegraphics[width=0.6\linewidth]{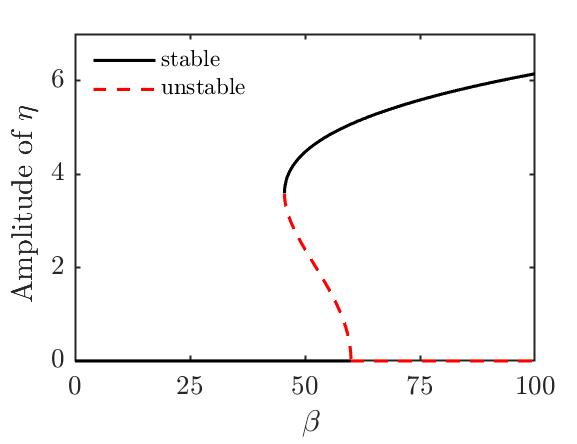}
\\
\end{minipage}
}
\end{figure}

\begin{figure}[h]
\centering
\includegraphics[width=0.7\linewidth]{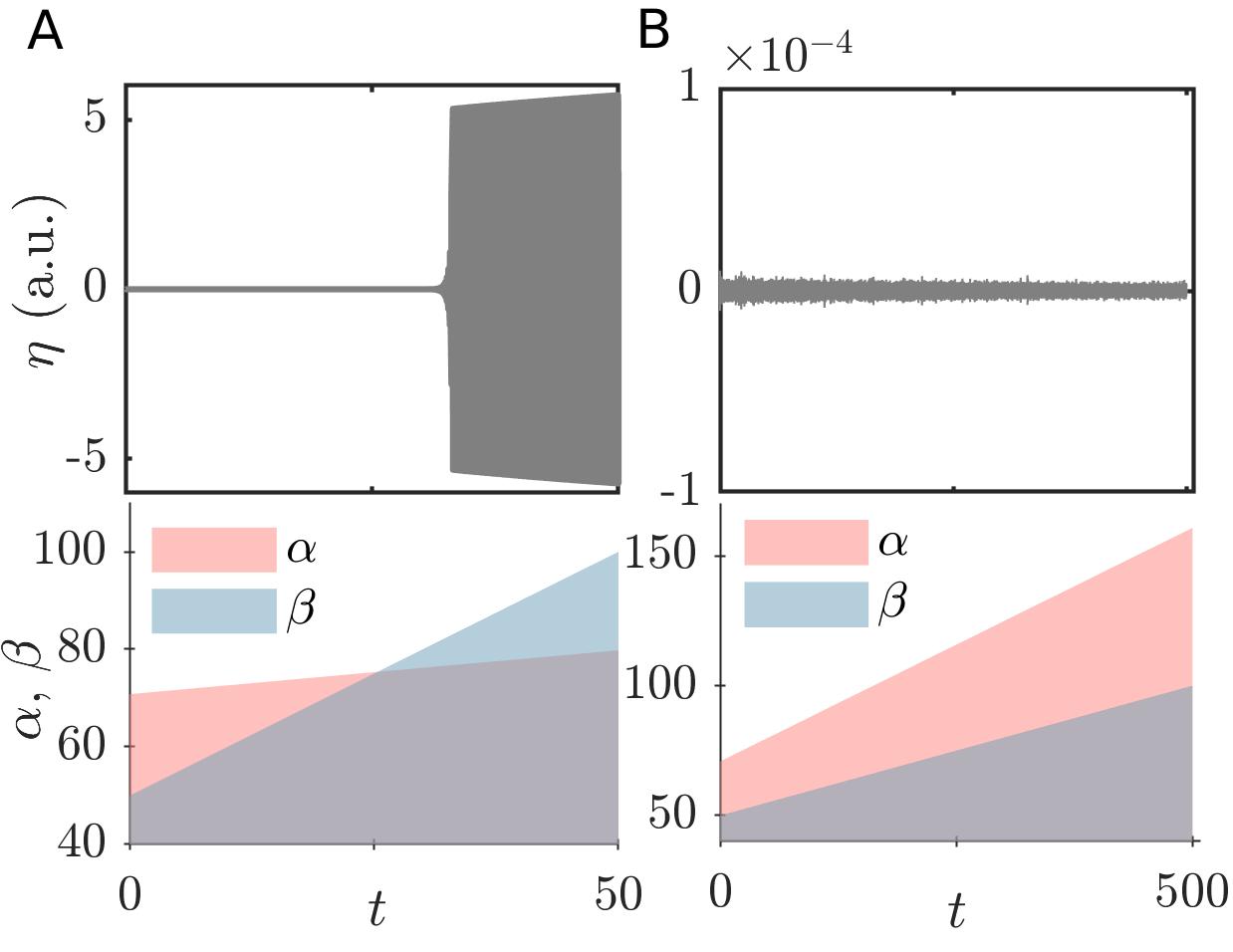}
\caption{Time series of $\eta$ obtained by solving the model described in Box~1 with continuously varying $\alpha$ and $\beta$. We vary $\alpha$ at a constant rate ($\dot{\alpha} = 0.18$) and $\beta$ at different rates: (A) fast ($\dot{\beta} = 1$) and (B) slow ($\dot{\beta} = 0.1$) rates. Here, a constant variation of $\alpha$ corresponds to a linearly increasing damping. Whenever the rate of change of driving ($\dot{\beta}$) is faster that the rate of variation of damping ($\dot{\alpha}$), we observe tipping ($\dot{\beta}> \dot{\alpha}$). High amplitude limit cycle oscillations occur when driving exceeds damping (A). No tipping is observed for $\dot{\beta}<\dot{\alpha}$ as driving never exceeds damping (B).}
\label{fig:fig1}
\end{figure}

The experimental system exhibits a transition from a state of low amplitude aperiodic oscillations to high amplitude periodic oscillations. We use a simple model where a similar transition is observed by varying two parameters independently to illustrate R-tipping. 
The nonlinear oscillator discussed in Box~1 exhibits a subcritical Hopf bifurcation from a state of low amplitude aperiodic fluctuations to a high amplitude limit cycle oscillations when the linear driving and damping are equal ($\beta=\alpha$). Generally, the effective driving, i.e., $\beta- \alpha$, is considered as the control parameter in such models \cite{noiray2017linear}. However, in reality, various processes contribute to driving and damping, and those processes occur independently at different rates. Hence, we consider such a scenario where $\beta$ and $\alpha$ vary at different rates. 
\begin{align}
     \beta = \beta_0 +\dot{\beta}t\\
    \alpha = \alpha_0+\dot{\alpha}t.
\end{align}
Motivated by the experiments, we choose $\alpha$ as a monotonically increasing function of time, varying at a constant rate ($\dot{\alpha}$) and $\beta$ is our control parameter. We vary $\beta$ at different rates ($\dot{\beta}$) - slower or faster compared to the rate of variation of $\alpha$. Figure~\ref{fig:fig1} shows two cases with faster and slower rates of $\beta$ compared to $\alpha$. When $\dot{\beta} > \dot{\alpha}$,
the transition to a limit cycle is observed for $\beta>\alpha$ (Fig.~\ref{fig:fig1}A). In the second case, wherein $\beta$ is varied at a relatively slower rate, the driving never exceeds damping, and the system does not transition to limit cycle oscillations (Fig.~\ref{fig:fig1}B).

\begin{figure}[h!]
\centering
\includegraphics[width=0.6\linewidth]{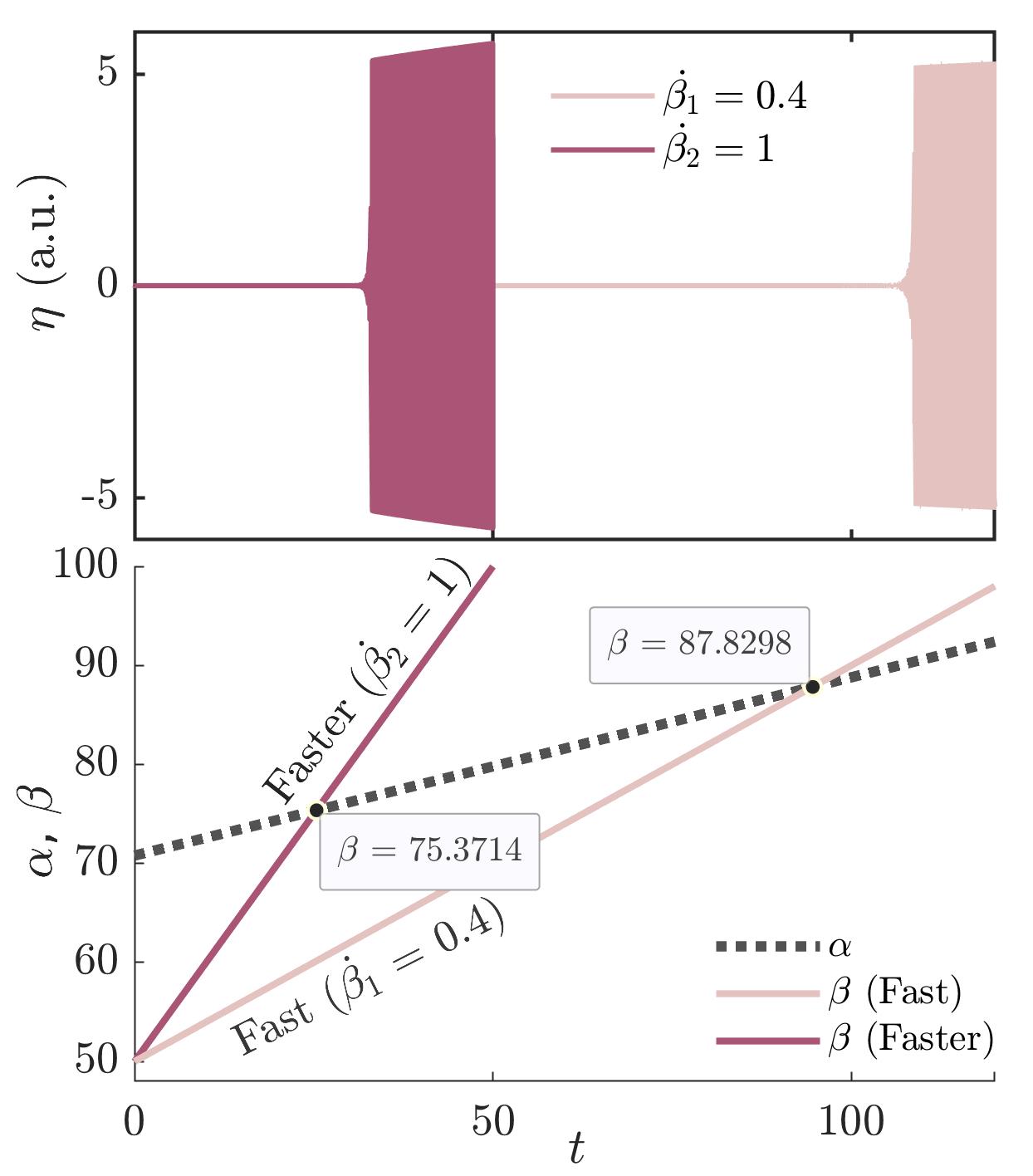}
\caption{Time series of $\eta$ and the corresponding variation of $\alpha$ and $\beta$ are shown for two different rates change of $\beta$. In both the cases, $\dot{\beta}>\dot{\alpha}$, i.e., $\beta$ is faster than $\alpha$. The onset of tipping occurs when driving crosses damping. However, as the damping varies in time, we need more driving to overcome damping for slow variation of driving. }
\label{fig:fig2}
\end{figure}

For the driving to match the damping, the variation of $\alpha$ and $\beta$ must intersect at some point, as shown in Fig.~\ref{fig:fig1}A. This happens only when $\beta$ varies faster than $\alpha$, as we start the system from a state of $\alpha>\beta$ (noisy fixed point state). In other words, the rate of change of $\beta$ should be greater than that of $\alpha$ for tipping to occur in the system ($\dot{\beta} > \dot{\alpha}$).
When $\dot{\beta} =\dot{\alpha}$, the trajectories of $\alpha$ and $\beta$ do not intersect. If $\dot{\beta} >\dot{\alpha}$ and they meet in a finite time within the duration of experiments, tipping occurs. Thus, there exists a critical rate of change of $\beta$ above which tipping occurs. However, it is difficult to find out such critical rate in experiments where $\dot{\beta} =\dot{\alpha}$. For $\dot{\beta}$ slightly greater than $\dot{\alpha}$, the driving intersects damping at very high value of driving (at a very large $t$), thereby making it hard to detect the correct critical rate in experiments. 

The Hopf bifurcation model discussed here is a representative case where competing parameters vary continuously at finite rates. The purpose of using the present model is not to fully describe the experimental thermoacoustic system; instead, it is to explain the mechanism in a simple general model. This result is in fact applicable to any type of transitions wherever there are two parameters with competing effects that vary in time.

One important aspect to keep in mind here is the rate-dependent tipping delay, which is well studied in the literature. Baer et al. \cite{baer1989slow} have discovered bifurcation delay when the control parameter is swept through the bifurcation point, and the same has been confirmed later by many others \cite{bonciolini2018experiments,park2011slow,berglund2000dynamic,majumdar2013transitions}. Such rate-dependent delay due to memory effects is commonly observed and is more prominent for faster rates. Hence tipping gets delayed from the bifurcation point for faster rates. Whenever there is a continuous variation of parameter with slow passage through bifurcation, there is always a rate-dependent delay in tipping from this point. However, the advancement due to a fast rate of change of parameter is substantially more than this rate-dependent delay in this case, because the driving crosses damping at a much lower value.

Next, we examine the dynamics for two different rates of change of $\beta$, i.e., $\dot{\beta_1}>\dot{\beta}_2>\dot{\alpha}$. We do not observe rate-dependent delay; in contrast, we observe advanced tipping for faster rates of variation of the control parameter ($\beta$). Figure~\ref{fig:fig2} shows the dynamics corresponding to $\dot{\beta}_1 =1$ and $\dot{\beta}_2 =0.4$. For $\dot{\beta}_1 =1$, the variation of $\beta$ is fast enough to cross $\alpha$ earlier before it grows to high magnitudes, whereas $\beta$ crosses $\alpha$ at a larger value (analogous to higher $Re$ in experiments) for the slow rate ($\dot{\beta} =0.4$). Thus, we observe tipping at lower values of $\beta$ as we vary $\beta$ faster. 

\begin{figure}[h!]
\centering
\includegraphics[width=0.8\linewidth]{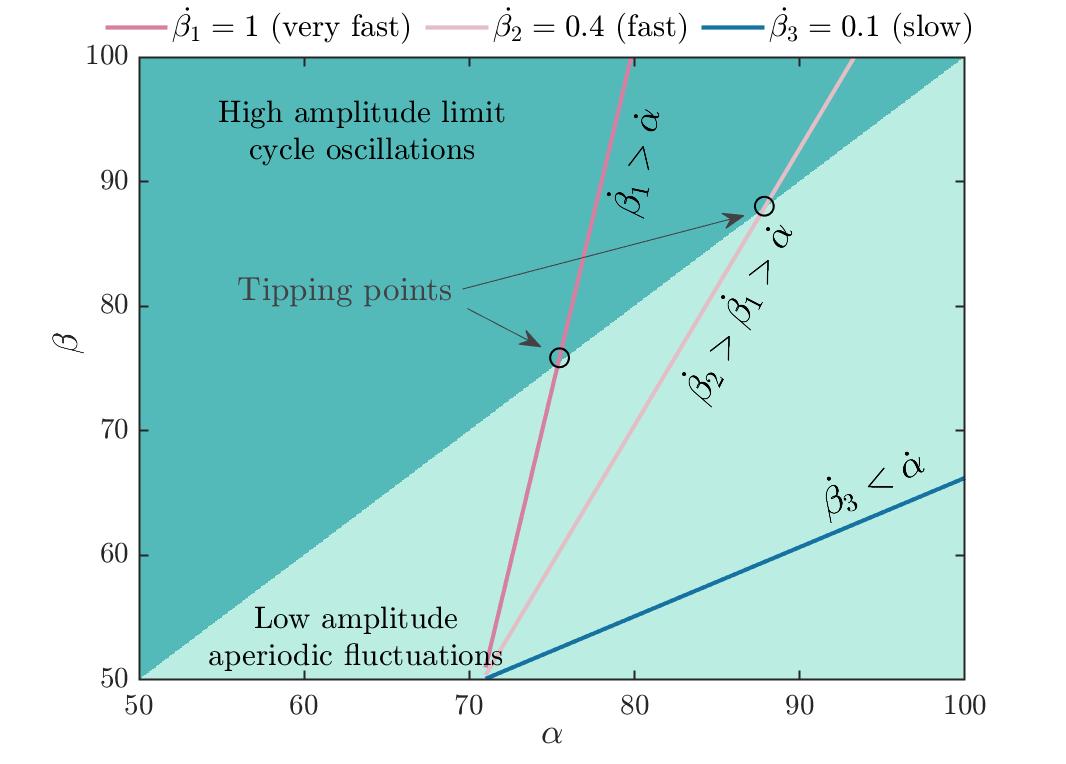}
\caption{The 2D bifurcation has two types of dynamics separated by the diagonal line, and the trajectories crossing this line undergo tipping. Above this line, driving is greater than damping resulting in limit cycle oscillations, and below the diagonal line, the dynamics consist of low amplitude aperiodic fluctuations. The trajectories crossing this line exhibit tipping. Relatively faster rates of change of $\beta$ cross at low values of $\alpha$. The slow rates of variation of $\beta$ that are slower than a critical rate do not cross this diagonal line even at very high values of driving. The variations in $\alpha$ and $\beta$ are analogous to variations in the process contributing to damping and driving in the experiments such as the wall temperature and $Re$.}
\label{fig:2d}
\end{figure}

Let us summarize the observations from the model in a 2D bifurcation plot (see Fig.~\ref{fig:2d}). In this model, bifurcation-induced tipping is observed when the linear damping is kept constant and driving is increased as the bifurcation parameter\cite{noiray2017linear}. At some value of driving, it matches damping and tipping occurs, irrespective of the rate at which the driving increases. As shown in the following figure, there is no tipping in the system for slow rates of variation of driving. The diagonal line in the $\alpha-\beta$ plane (see Fig.~\ref{fig:2d}) separates the two dynamical regimes, and the line represents the bifurcation points.
Whenever $\alpha$ varies independently at a finite rate, slow variations of $\beta$ (relative to the rate of change of $\alpha$) do not result in tipping; only faster rates induce tipping. In other words, there is tipping whenever $\beta$  crosses $\alpha$, however, it is possible only with faster variations of $\beta$ above a critical rate. The critical rate here is the rate of variation of $\alpha$. 

We cannot incur bifurcation tipping in this particular configuration of the experimental system unless there is a way to keep the wall temperature constant. The rates of variation of air flow rate should be classified as fast or slow compared to the rate of increase of damping. Such an explicit comparison is difficult in experiments. However, from the model, we understand that only faster rates will result in tipping. So, we defined the rates which resulted in tipping as fast rates and others as slow.

\section*{Discussion and conclusions}

In this work, we show the possibility of rate-induced tipping in a turbulent thermoacoustic system. When we vary the Reynolds number continuously at a slow rate, the system remains in the same dynamical state and does not exhibit any tipping. In contrast, if the Reynolds number is varied fast enough, we observe a transition to a state of thermoacoustic instability with high amplitude limit cycle oscillations. In order to understand the mechanism behind this rate-induced tipping in the context of transition to thermoacoustic instability, we explore the variations of another system parameter, the wall temperature, that affects acoustic damping. 

We discover that the wall temperature varies continuously at a slow rate, i.e., a parameter which is not in our control  varies simultaneously at a different rate. 
Experiments conducted at faster rates of change of $Re$ last for shorter duration and the increase in wall temperature is smaller. In contrast, the wall temperature increases gradually to very high values for experiments conducted at slower rates of change of $Re$, which last for a longer duration.
Then, we measure the acoustic damping by estimating the decay rate of oscillations in the system for different wall temperatures and discover that the damping increases linearly with increasing temperature. Thus, we infer that the fast variation of parameters could take the system to a different set of parameters where the damping is very low and can be excited to a high amplitude limit cycle that was otherwise inaccessible. 

The increase in $Re$ drives the system towards thermoacoustic instability, whereas the increasing wall temperature increases damping and, in turn, reduces the effective driving. Here, we find a mechanism of R-tipping, where two parameters that have the opposite effect on tipping compete to determine the dynamics. Motivated from the experimental observations, we use a simple system-independent model of Hopf bifurcation to illustrate this phenomenon. When we vary a control parameter which drives the system towards tipping and there exists another parameter that has a negative effect on transitions, the control parameter has to be varied at very high rates to achieve tipping. This mechanism and explanation for R-tipping are very general and system independent. To date, such a type of R-tipping has not been explained in any experimental systems. As R-tipping in real-world systems is extremely dangerous, understanding the mechanism of R-tipping would enable better predictions of R-tipping.

In the context of thermoacoustic systems, while estimating the stability map of combustors, the rate of change of the control parameter, for example, engine throttling rate, should be considered as a parameter to define the safe operating boundary of a thermoacoustic system. To understand the real stability margin, the engine has to be subjected to highest expected throttling rate to determine whether there is a critical rate above which thermoacoustic instability is present in the system. Further, we note that the mechanism by Tony et al.\cite{tony2017experimental} considers the variation of a single parameter in time, and it is achieved by preconditioning. Ours is a new mechanism as we consider variations of two parameters that have opposite effects on tipping, and it is the competition between the effects of these parameters that determines the tipping. 

The applicability of our mechanism of R-tipping is not restricted to thermoacoustic systems. In fact, it can be observed in any system where two parameters vary simultaneously, given that each has the opposite effect on the driving mechanism of a critical transition. Then, the competition between these two variables determines whether and when the tipping will occur. For instance, this mechanism can explain the potential climate tipping point, the `compost bomb instability'\cite{luke2011soil}. The compost bomb instability occurs above a specific rate of global warming when heat is generated in the soil faster than it can escape to the atmosphere. Here, the generation of heat in the soil and its escape to the atmosphere are two phenomena with opposite effects on tipping. Depending on the relative rate of heat generation and its escape, compost bomb instability occurs. Our mechanism correctly explains the observation of the critical rate in compost bomb instability depending on the relative rate of heat generation and its escape. 
Wieczorek et al. \cite{wieczorek2011excitability} have studied R-tipping to compost bomb instability in an analytical framework and derived conditions for the critical rate and excitability threshold. However, they have viewed it from a different perspective of excitable systems. 
This mechanism of R-tipping is general and can be observed in systems where two parameters vary simultaneously, given that each has the opposite effect on the driving mechanism of tipping. Then, the competition between the two variables determines the dynamics.

\section*{Materials and methods} 
\subsection*{Details of measurements from the experiments}

The combustion chamber is 800 mm long. The air flow enters through the inlet to the plenum chamber. The fuel (liquefied petroleum gas (LPG): 60\% butane and 40\% propane) is injected upstream of the combustion chamber. We ignite the partially premixed reactant mixture using a spark plug. We use a fixed vane swirler of diameter $d = 40$ mm for flame holding. The swirler has 8 vanes, with a vane angle of $40^{\circ}$ with respect to the longitudinal axis. Once the flame is stabilized in the combustor, we vary the mass flow rate of air, which, in turn, varies the Reynolds number ($Re$). 

In the present study, we vary $Re$ continuously by controlling the mass flow rate of air, $\dot{m}_a$, as a linearly increasing function of time. The mass flow rate of fuel, $\dot{m}_f$ is kept constant at 0.76 g/s, and $\dot{m}_a$ is increased from 7.76 g/s to 13.07 g/s at different rates, and $Re$ varies from $1.3\times10^4$ to $2.2\times10^4$. The flow rates of air and fuel are controlled using mass flow controllers (Alicat, MCR series) with an uncertainty of $\pm$(0.8 $\%$ of reading + 0.2 $\%$ of full scale). The corresponding maximum uncertainty in $Re$ is $\pm 2.5 \%$. The $Re$ varies from $1.3\times10^4$ to $2.2\times10^4$. The corresponding variation in equivalence ratio is from 0.99 to 0.59. The equivalence ratio is defined as $\phi = \frac{{(\dot m_f/\dot m_a)}_{actual}}{{(\dot m_f/\dot m_a)}_{stoichiometry}}$, where $\dot m_f$ and $\dot m_a$ are the mass flow rates of fuel and air, respectively. 

We measure the pressure fluctuations inside the combustor (at a sampling rate of 4 kHz) using piezoelectric pressure transducer (PCB103B02) mounted at a distance of 360 mm from the swirler. The sensitivity of the transducers is 217.5 mV/kPa, and the maximum uncertainty is $\pm 0.15$ Pa. A K-type thermocouple is used to measure the wall temperature of the combustion chamber close to the swirler at a distance of 90 mm from the backward facing step. The pressure signals are acquired using a 16-bit A/D card (NI-6343), and the temperature data are acquired by a 24-bit A/D card (NI-9211) at a sampling rate of 4 Hz.

\subsection*{Measuring damping at high temperatures} 
To understand the role of wall temperature in R-tipping, we have to determine how a change in wall temperature contributes to the damping/driving in the system. 
At room temperatures, without any inlet flow, we can measure damping by exciting the natural modes of the combustor duct using an externally driven speaker. A sinusoidal pressure perturbation is generated, the speaker is abruptly switched off, and the decay rate of periodic oscillations is estimated. We extract the envelope of amplitude and fit a line in the semi-logarithmic plot of the envelope to estimate the decay rate. The decay of amplitude of the signal is fitted with an exponential ($e^{-\zeta t}$) function. 

\begin{figure}[h!]
\centering
\includegraphics[width=0.4\linewidth]{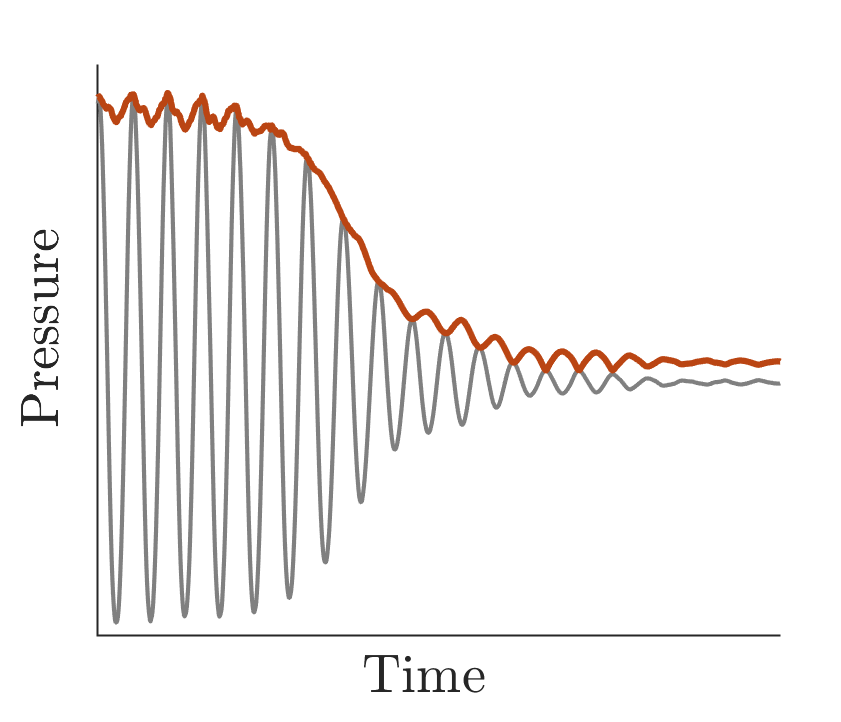}
\caption{A representative acoustic pressure signal showing the decay of oscillations when the fuel supply is cut abruptly. We extract the amplitude envelope of oscillations and fit a line in the semi-log plot to calculate the decay rate.
}
\label{fig:fig7}
\end{figure}

It is hard to measure damping at high temperatures by using externally driven speakers, as done for room temperature measurements. Therefore, we use a different method, where we establish the system in the state of thermoacoustic instability and abruptly cut off the fuel supply and stop the combustion. We start the experiments by keeping all the operating conditions but the wall temperature the same, and then vary $Re$ to reach the state of thermoacoustic instability. The wall temperature is recorded throughout the experiment, and the specific value is noted at the instant when the flow of fuel is cut.

\bibliography{main}

\bibliographystyle{Science}

\section*{Acknowledgments}
We thank S. Anand, S. Thilagaraj, G. Sudha and M. Ragunathan for their help during the experiments. R.I.S. acknowledges the Science and Engineering Research Board (SERB) of the Department of Science and Technology for funding under grant no: CRG/2020/003051. I.P. acknowledges the research assistantship from the Ministry of Human Resource Development, India and Indian Institute of Technology Madras.

\section*{Author contributions}
R.I.S. conceived the idea and designed the study. I.P. and P.R.M. conducted the experiments with inputs from R.I.S. I.P. and R.I.S. performed data analysis and modelling. All authors discussed the results and participated in writing the manuscript.
\section*{Competing interests} The authors declare that they have no competing interests.
\section*{Data availability}
All the data presented in this paper are available from the corresponding author on reasonable request.


\end{document}